\documentclass[10pt,journal,compsoc]{IEEEtran}
\ifCLASSOPTIONcompsoc
  \usepackage[nocompress]{cite}
\else
  \usepackage{cite}
\fi
\ifCLASSINFOpdf
  \usepackage[pdftex]{graphicx}
  \graphicspath{{figures/}{pictures/}{images/}{./}}
  \DeclareGraphicsExtensions{.pdf,.jpeg,.png}
\else
\fi
\usepackage{amsmath}
\usepackage{algorithmic}
\usepackage{fixltx2e}

\usepackage{stfloats}
\usepackage{url}

\begin{document}
%
\title{Improving the Projection of Global Structures in Data through Spanning Trees}
%
%
%
%

\author{Daniel Alcaide, and Jan Aerts 
\IEEEcompsocitemizethanks{
\IEEEcompsocthanksitem  Daniel Alcaide is with ESAT/STADIUS, Faculty of Engineering,KU Leuven and is the corresponding author. 
E-mail: daniel.alcaide@kuleuven.be
\IEEEcompsocthanksitem Jan Aerts is with I-BioStat and Data Science Institute, UHasselt, and ESAT/STADIUS, Faculty of Engineering, KU Leuven. \protect\\
E-mail: jan.aerts@uhasselt.be
}
}

\IEEEtitleabstractindextext{%
\begin{abstract}
The connection of edges in a graph generates a structure that is independent of a coordinate system. This visual metaphor allows creating a more flexible representation of data than a two-dimensional scatterplot. In this work, we present STAD (Spanning Trees as Approximation of Data), a dimensionality reduction method to approximate the high-dimensional structure into a graph with or without formulating prior hypotheses. STAD generates an abstract representation of high-dimensional data by giving each data point a location in a graph which preserves the distances in the original high-dimensional space. The STAD graph is built upon the Minimum Spanning Tree (MST) to which new edges are added until the correlation between the distances from the graph and the original dataset is maximized. Additionally, STAD supports the inclusion of additional functions to focus the exploration and allow the analysis of data from new perspectives, emphasizing traits in data which otherwise would remain hidden. We demonstrate the effectiveness of our method by applying it to two real-world datasets: traffic density in Barcelona and temporal measurements of air quality in Castile and Le\'on in Spain.
\end{abstract}


}

\maketitle

\IEEEdisplaynontitleabstractindextext

%
\IEEEpeerreviewmaketitle

\IEEEraisesectionheading{\section{Introduction}\label{sec:introduction}}

%
%
%
%
\IEEEPARstart{D}{ata} visualization is extensively used to reveal patterns and structures in data. The display of high-dimensional datasets concerning point clouds with a high number of attributes continues to be a relevant research field due to the wide range of applications. The choice of an informative visualization technique depends not only on the characteristics of the data but also on the tasks to be performed. For example, a visualization to analyze the evolution of a high-dimensional time series requires a different approach than projecting a document corpus. While both aim to represent the data in a limited number of dimensions, the first emphasizes the progressive and continuous changes that occur in time and the second aims to find differences between groups of documents.

Dimensionality reduction techniques allow for embedding high-dimensional data into a plot with two or three axes. These solutions provide visual scalability advantage over classical scatterplot matrices and parallel coordinates \cite{munzner2014visualization}. However, the low-dimensional transformations rely on assumptions and parameterizations which can compromise part of the original information. The most recent methods such t-SNE \cite{maaten2008visualizing} or UMAP \cite{mcinnes2018umap} are effective in identifying similar elements and projecting them separated from other groups. These projections focus on preserving the closest neighbors rather than preserving all distances between the points which can cause an incomplete mapping of the dataset in the lower space \cite{schubert2017intrinsic}.

On the other hand, Topological Data Analysis (TDA) aims to deduce and recognize geometric structures from underlying data by means of connecting elements in a graph. The combination of a scalar function with the original source allows exploration of data from different perspectives highlighting information which otherwise would be hidden. Unlike dimension reduction techniques the reconstruction of topology may not be faithful to the original data geometry, but they preserve the continuity between data shapes. Although TDA has demonstrated remarkable results in specialized studies [\citenum{lum2013extracting}, \citenum{nielson2015topological}, \citenum{lakshmikanth2017mass}], it relies on data summaries (e.g. clusters) instead of individual data points and therefore limits the resolution of exploration phase. In addition, the cornerstone of TDA is clustering and appropriate lens which precludes hypothesis-free data exploration.

\begin{figure*}
  \centering
  \includegraphics[width=\linewidth]{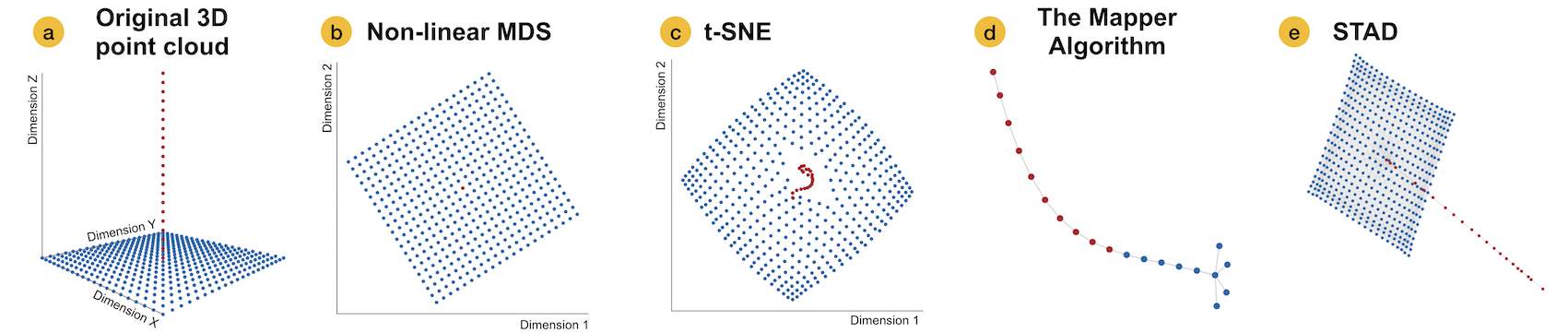}
  \caption{
  Comparison of methods to visualize a three-dimensional point cloud.   
  (a) Three-dimensional point cloud composed of a two-dimensional grid and orthogonal vector. 
  (b) Non-linear multidimensional scaling projection. The influence of the two-dimensional grid avoids the representation of the orthogonal vector located in the middle of the grid.
  (c) t-SNE projection allows the visualization of the grid and the vector. However, the projection distorts the signal losing the connection between both patterns.
  (d) The Mapper algorithm using the L-infinity centrality as the lens. The aggregation reduces the dataset, but the influence of lenses drives the projection.
  (e) STAD projection without additional filters. The flexibility of the network structure allows the projections of the three-dimensional dataset, although it can be extrapolated to higher dimensionality.
}
	\label{fig:stad_dim_red}
\end{figure*}

In this paper we present STAD, a parameter-free dimensionality reduction method which transforms the high-dimensional data into a graph highlighting the underlying structure. The projection into a graph provides a higher degree of flexibility to represent the interdependencies between points than coordinate mapping techniques (Fig. \ref{fig:stad_dim_red}). Furthermore, STAD allows for incorporating additional functions which can intensify the specific signals adding new perspectives to the exploratory analysis. Additionally, STAD networks generate a representation of data without aggregation, i.e., STAD encodes the original data points as vertices in the graph which provides a high resolution of the data. 

This paper is organized as follows. In section 2 we give an overview of related work in the detection of structure in data through dimension reduction and exploratory techniques using graph representations. Section 3 describes the proposed methodology, followed by section 4, in which we present two case studies applying STAD. The approach is discussed in section 5, and finally, section 6 covers conclusions and possible directions for future work.

\section{Related Work}

The exploration of high-dimensional data has been presented in different areas of research in information visualization, data mining,  and machine learning \cite{liu2017visualizing}. We review the related work in these areas below, more specifically the topics of dimension reduction, visualization techniques and TDA.

\subsection{Dimension reduction and embeddings}

Dimensionality reduction techniques transform a high-dimensional space to a low dimensional one. Considering as input the $N^2$ data points of a pairwise distance matrix, the methods to visualize the global structure falls in the multidimensional scaling (MDS) class \cite{mukherjee2018multidimensional}. Torgerson Scaling \cite{torgerson1958theory}, a particular case of PCA, finds a linear and orthogonal transformation of data revealing the most informative view without modifying the local and global relationship between elements. More versatile approaches are non-linear metric MDS (NMDS) \cite{kruskal1964nonmetric} and Sammon mapping \cite{sammon1969nonlinear}, which overcome the linearity limitation introducing an iterative approach to match the distances between the original and projected space by minimizing the error between both matrices. The difference between Sammon mapping and non-linear MDS is that Sammon normalizes the original distance to emphasize small differences. The iterative MDS approach has been the basis for other models with improved versions of the loss functions \cite{saeed2018survey}. 

The Isomap algorithm \cite{balasubramanian2002isomap} is also based on the iterative MDS model, but using geodesic distances instead of Euclidean distance. The algorithm defines a neighboring region based on a parameter $\varepsilon$ given by the user. Once the neighbors are defined, the low-dimensional embedding is generated in a similar fashion to iterative MDS. This method eliminates the need of estimating distances between widely separated elements.

The underlying idea of associating the pairwise distances between the original space and a projected space is also employed in the STAD method. The difference with MDS techniques is that the projection takes place into a graph and more precisely in the path described by nodes and edges. The change in spatialization from a fixed coordinate system to a free-dimensional space provides a more flexible visual technique to represent the information as compared in Fig. \ref{fig:stad_dim_red}. Instead of mapping the position in a  lower space, the STAD graph aims to visualize the relationships between the data elements.

Beyond MDS techniques, more recent dimensionality reduction techniques such as t-SNE \cite{maaten2008visualizing}, LargeVis \cite{tang2016visualizing} and UMAP \cite{mcinnes2018umap} improve the projection of low-dimensional space by intensifying the detection of nearest neighbors. These techniques assume an intrinsic probability distribution which smooth the projection in the low-dimensional space improving the detection of local patterns distorting the global one. However, these approaches tend to increase the division among neighborhoods, which are beneficial for identifying local clusters but contrary to the accurate identification of global trends or continuous patterns \cite{schubert2017intrinsic}. 

\subsection{Exploration of data through network structures}

A number of earlier research projects used the network metaphor to facilitate the understanding of multidimensional data (e.g., [\citenum{demiralp2013invis}, \citenum{alcaide2018mclean}, \citenum{stuetzle2003estimating}, \citenum{janicke2008brushing}]). All these techniques depend on a parameter which determines the elements connectivity. The exploratory system presented by J\"{a}nicke et al. \cite{janicke2008brushing} employs the the Minimum Spanning Tree (MST) to define the minimal structure of the data. Additional edges are added to establish a more consistent data structure using a force-directed graph layout. STAD uses the same concept of adding edges on top of a MST to draw the data shape but the number of edges are automatically selected through a minimization process. The structures presented in STAD generate a more accurate representation of the original high-dimensional space by providing not only the structural shape but also an intuition of the density through the interconnection of nodes in a region of the graph.

Topology studies the global structure of a data from a geometrical perspective providing an informative summary. Topological Data analysis is the general term used for a collection of particular methods to analyze high-dimensional datasets \cite{munch2017user}. Graph representations are commonly used to illustrate the underlying structure of data, but nodes are aggregations of points rather than individual elements. Under this umbrella, data skeletonization is an important shape descriptor from a disconnected point cloud [\citenum{zomorodian2010fast}, \citenum{kurlin2015one}, \citenum{rieck2014structural}]. The selection of a proper skeleton is defined by the representation which shows the most persistent features. The stability of topological features is visualized in a so-called barcode \cite{edelsbrunner2008persistent} and analyzed to identify suitable parameterization [\citenum{wasserman2018topological}, \citenum{ghrist2008barcodes}]. Other TDA methods rely on scalar functions to guide the summarization of high-dimensional data such as Morse-Smale Complexes (MSC) [\citenum{gerber2010visual}, \citenum{shivashankar2012parallel}, \citenum{gunther2012efficient}], Reeb graphs \cite{biasotti2008reeb} and the Mapper algorithm introduced by Singh et al. \cite{singh2007topological}. While these functions are defined to determine the continuous space of a manifold, the functions in STAD modify the projections of distances by controlling the connections between nodes. However, since the same functions can be applied in both structures, they can be similar. In addition the evaluation criteria in STAD relies on the association between the graph representation and the original space which differ from the geometrical persistent implications of TDA.

\section{Methodology}

Network visualization representations are projections of data expressed independently from the coordinate system; the visual structure connects elements according to their relationship and not their location. We extend the concept to represent the similarity (distance) between two nodes as the path described by the edges in the network. Once a similarity metric is chosen, a distance matrix $D_{X}$ containing the pairwise distances between all elements can be defined. The distance matrix can be considered a complete weighted graph $G_{X}$, where the indices of the matrix represent the vertices of the graph and the edge weights the distance between any two elements. 

STAD proposes a new method to generate the structure of data by transforming the distance matrix into a graph. This method converts the complete graph $G_{X}$ into a non-complete unit-distance graph $U$ (i.e., all edges in the network have the same length of 1) where the distance between datapoints is reflected in the length of the shortest path between them. That is, the distance between two distant points is build from the neighborhoods of other nodes. A STAD network forms a single connected component, and it ensures the path for any combination of vertices exists. The information presented in the STAD networks is an approximation of the original complete weighted graph due to discretizing the distances in unitary segments. The number of edges in the unit-distance graph controls the shape of the data, and a final graph can vary from the minimum spanning tree to the complete graph. The STAD procedure selects the number of edges automatically by maximizing the correlation between the weighted distance matrix $D_{X}$ and the unit-distance graph.

In section 3.1, we describe the details of the STAD algorithm and illustrate the steps with a simulated example. In section 3.2, we present an extension of STAD to amplify signals in data through the addition of filters. 

\begin{figure}[!]
  \centering
  \includegraphics[width=\columnwidth]{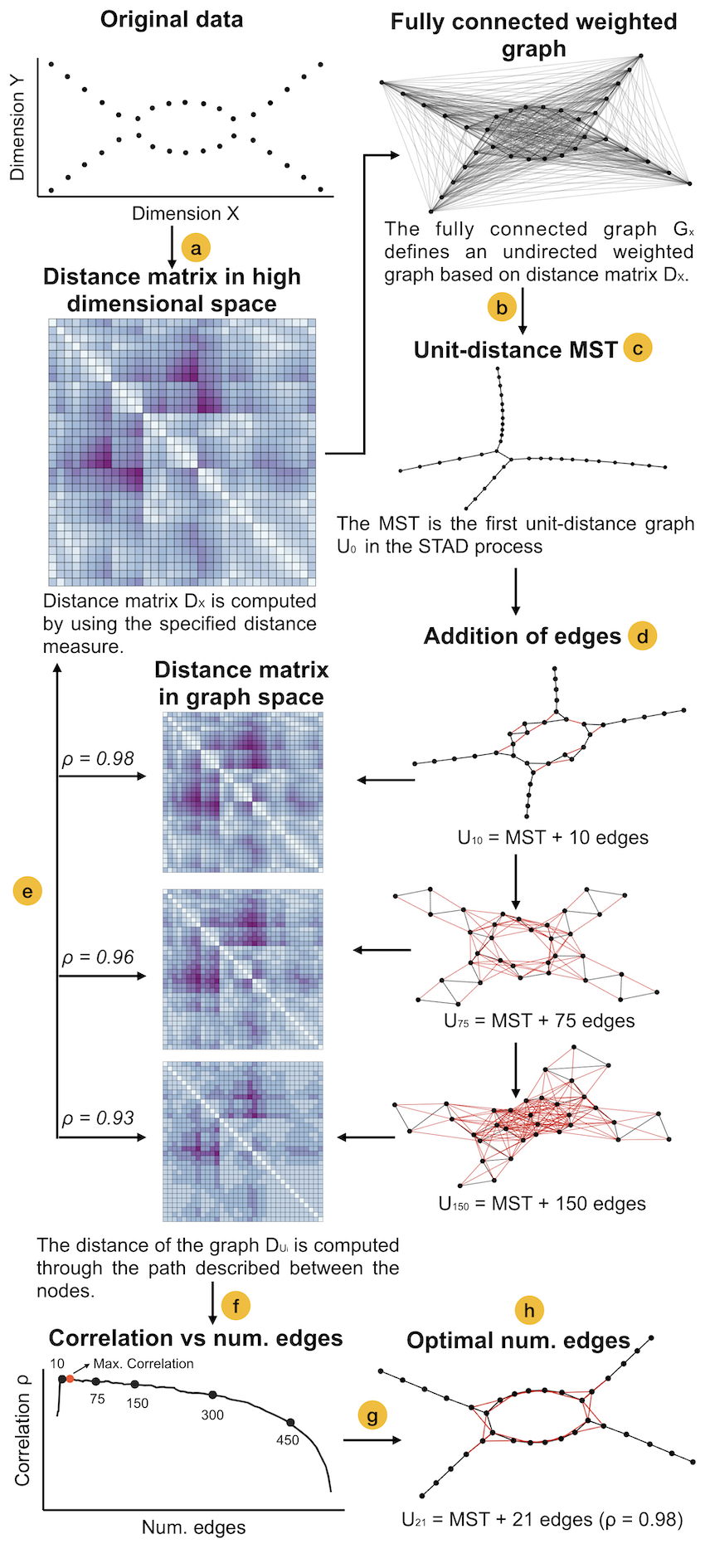}
  \caption{STAD base algorithm illustrated in eight steps: 
  (a) Create distance matrix $D_{X}$: From the point cloud and a defined distance metric, the pairwise distances between all elements are computed. The complete graph $G_{X}$ is derived from $D_{X}$ by encoding the distances as weights in the network.
  (b) Build MST from complete weighted graph: The MST is computed from the complete weighted graph $G_{X}$. 
  (c) Convert MST to unit-distance graph: The weights from the MST are removed and the path is used as a new measure of distance.
  (d) Add edges into unit-distance graph: Edges are sorted and added to the graph in sequential order.
  (e) Evaluate the objective function: The correlation is computed between the distances from the unit-distance graph and the original distance matrix.
  (f) Visualize the relationship between correlation and the number of edges.
  (g) Identify optimal number of edges: The optimal is found at iteration with maximum correlation.
  (h) Create a node-link diagram: Original distances are added as weights and a proper visual layout is chosen.
}
	\label{fig:stad_base}
\end{figure}

\subsection{STAD base algorithm}

The STAD algorithm can be split in eight sequential steps (Fig. \ref{fig:stad_base}): create the distance matrix, build an MST from the complete weighted graph, convert the MST to the unit-distance-graph, add edges to the unit-distance graph, evaluate the objective function, visualize the relationship between correlation and the number of edges, identify the optimal number of edges automatically and create a node-link diagram of the final network.

\subsubsection{Create distance matrix in original high-dimensional space}
\label{stad_base_step_1}

Let $X$ be a space with $n$ elements and $m$ dimensions in ${\rm I\!R}^m$, and a metric exists which determines all pairwise distances $d_{ij}$ with $1 \leq i$, $j \leq n$. The distance matrix $D_{X}$ is the squared matrix with size $n\:x\:n$ containing all the distances $d_{ij}$. This distance matrix $D_{X}$ is the only required element to generate a STAD network. From our perspective, the matrix $D_{X}$ is understood as an undirected, weighted and fully connected graph $G_{X}$ with $n$ vertices and $\frac{n^{2} - n}{2}$ edges where each edge $e_{ij}$ has a weight of value $d_{ij}$. Fig. \ref{fig:stad_base}a illustrates the distance matrix creation from a point cloud and the representation of the fully connected graph. The similarity between each pair of elements is projected as edges in the graph. Notice that edges in $G_{X}$ are undirected due to the symmetric property of the matrix and the diagonal elements are omitted in the graph.

\subsubsection{Build MST from complete weighted graph}
\label{stad_base_step_2}

Next, a minimum spanning tree (MST) is computed for $G_{X}$. The MST is a subset of $n-1$ edges that connects all vertices without loops, with a minimal sum of edge weights. Fig. \ref{fig:stad_base}b shows the MST network for the complete graph $G_{X}$. Note that the MST may not be unique and alternative combinations of edges can produce the same result.

\subsubsection{Convert MST to unit-distance graph}

The MST is the first unit-distance graph $U_{0}$ considered in STAD (Fig. \ref{fig:stad_base}c) and the addition of edges will improve the association between $U$ and $G_{X}$. By transforming the complete graph $G_{X}$ into a unit-distance graph $U$, we reduce the graph dimension of the original into a two-dimensional graph formalized by Erd\H os \cite{erdos1965dimension}.

\subsubsection{Add edges into unit-distance graph}
\label{add_edges}

The addition of new edges to the unit-distance graph $U$ produces a new graph $U_{i}$, where $i$ is the number of edges included in addition to the MST.

First, edges are sorted by weight to define the order in which they are added. For instance, $e_{i-1} < e_{i}$ means that the weight of $e_{i-1}$ is smaller than that of $e_{i}$, or that the datapoints referred to in $e_{i-1}$ are closer together in the original space than those referred to in $e_{i}$. Then the edges are added into the network $U_{i}$ following a cumulative process so that if $e_{1}<e_{2}<e_{3}<...<e_{n-1}<e_{n}$ then $U_{1}\subset U_{2} \subset U_{2} \subset ... \subset U_{n-1} \subset U_{n}$ where $U_{i}$ is the unit-distance graph with the sequence of edges $e_{1}$, $e_{2}$, $e_{3}$, ..., $e_{i}$ from $G_{X}$. Although edges are sorted and added into the unit-distance graph by their weight, all edges in $U$ itself are unweighted. Fig. \ref{fig:stad_base}d shows three examples of unitary graphs to give an intuition into how the network evolves by adding new edges in addition to the MST. The number of possible unitary graphs $U$ depends on the number of data elements in the space $X$, so that $\frac{n(n-1)}{2}+1$.

\subsubsection{Evaluate the objective function: calculate correlation  between distances in original space and those in graph space}

From the unit-distance graph $U_{i}$, the computation of the shortest path for every pair of vertices produces a squared matrix $D_{U_{i}}$ with size $n\:x\:n$ comparable to the distance matrix $D_{X}$ for the original space $X$. The Pearson correlation is used to measure the agreement between $D_{U_{i}}$ and $D_{X}$. Contrary to the statistics absolute error, correlation is invariant under different scalings and takes a known range within -1 and 1 \cite{lee1988thirteen}. The correlation between the two matrices during the STAD process is limited to the range from 0 to 1 because the distances projected in $D_{U}$ follow a similar mapping, i.e. long distances in $D_{X}$ are projected as long distances in $D_{U}$. This finite range provides an intuition of the algorithm performance and a comparable benchmark between iterations at all levels of data. Fig. \ref{fig:stad_base}e exemplifies the changes in the distance matrix $D_{U_{i}}$ for different values of $i$ together with association value with the original distance matrix $D_{X}$.   

\subsubsection{Optional: Visualize the relationship between correlation and the number of edges}

The evaluation of the correlation for consecutive values of $i$ describes a quasi-convex function. The influence of an edge addition at the beginning of the sequence (i.e., at low values of $i$) has a large effect on the correlation with distance matrix $D_{U}$. The evaluation of correlation at this stage may fluctuate when $U$ contains few edges but describes a convex curve when the amount of edges is big enough. The number of edges needed to reach maximum correlation depends on the size and nature of the data.
Intuitively, the association between $D_{X}$ and $D_{U}$ is related to the concept of persistence of clustering solutions \cite{DBLP:journals/corr/abs-1811-00102} , i.e., if the shape is persistent along the edges, the computed correlation will be consistently similar. Fig. \ref{fig:stad_base}f shows the correlation curve for multiple evaluations. The correlation value initially increases by adding edges on top of the MST, reaching its maximum quickly. After the maximum, there is a constant decrease in the correlation when adding more edges demonstrating that these additional edges degrade the projection of data.

\subsubsection{Identify the optimal number of edges automatically}

STAD uses simulated annealing (SA) to estimate the optimal $\hat{U_{i}}$ which maximizes the correlation and provides a representative data projection of $X$. SA approximates the combination of links which maximize the correlation between $D_{U_{i}}$ and $D_{X}$. This heuristic is a stochastic process and estimates the global optimum by exploring the discrete space of edges (Fig. \ref{fig:stad_base}g). The SA candidate generator reduces failures on non-convexity regions produced by the correlation, mainly when the graph is composed by few edges. Note that as the resulting network is ultimately explored visually and structural features are kept along a range of edges, the difference between a global maximum and an approximation thereof does not imply noticeable deviations in STAD. The start and end of these ranges are only identifiable after the evaluation of all links. Fortunately, these do not all have to be computed, as the most important characteristics are also the most easily detectable.

The time complexity of calculating $D_{U_{i}}$ is $O(\left|V\right| + \left|E \right|)$ as described in the Breadth-first search algorithm definition in \cite{skiena2012weighted}; where $\left|V\right|$ is the number of vertices and $\left|E \right|$ is the number of edges in the graph. Although the number of vertices is fixed, the number of edges evaluated at iteration i influences the running time and varies between $O(\left|V- 1\right|)$ when the MST is evaluated and $O(\left|V\right|^2)$ when the graph is complete.

\subsubsection{Create node-link diagram of final network}
\label{node_link_diagram}

Networks require a graphical convention to be visualized. Generally, they are drawn as node-links representations projected in the two-dimensional plane. Although the STAD methodology generates an unweighted graph (unit-distance graph), we include the distances from the original metric $D_{X}$ in the final node-link diagram as this improves the visual representation. Fig. \ref{fig:stad_base}h shows the resulting network including the distances as weights in the edges. 
Note that the STAD graph is independent of the graph drawing algorithm used, the focus is on the identification of signals described by the connections of elements rather than the coordinates of nodes. However, graph drawings which minimize the number of crossings and place together small edge weight are appropriate to detect data structures, e.g., ForceAtlas2 \cite{jacomy2014forceatlas2} and Kamada-Kawai \cite{kamada1989algorithm}. 

\begin{figure*}
  \centering
  \includegraphics[width=\linewidth]{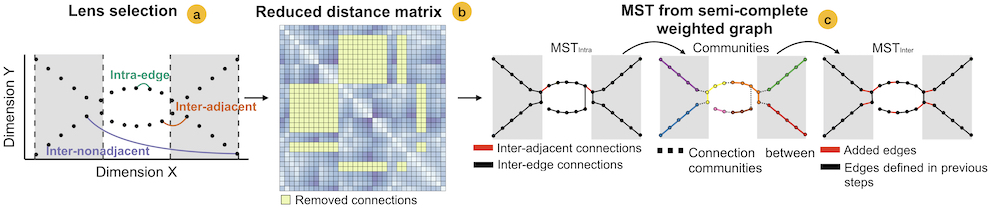}
  \caption{
  STAD algorithm extension for the integration of filters which substitutes the first two steps of the base algorithm: 
  (a) Define the filter: The figure illustrates the discretization of a real dimension $X$ in three intervals ($r=3$).
  (b) Create reduced distance matrix based on the filter: Inter-nonadjacent connections are omitted in the reduced distance matrix $D_{X}^{*}$.
  (c) Build the MST from this semi-complete weighted graph. This alternative MST is calculated in three steps. 1) MSTs are computed inside each filter index with intra-edges, inter-adjacent edges are added afterwards. 2) Intra-edges are validated through community detection. All inter-adjacent edges and intra-edges connecting different communities are removed. Intra-edges belonging to the same community are fixed and become part of the final MST. 3) Additional edges connect nodes in different connected components, thereby creating a single connected component.
}
	\label{fig:stad_lens}
\end{figure*}

\subsection{Filters in STAD}

The transformation of datapoints and weights from a complete graph into a unit-distance graph representation can hide part of the information, and we can expect that not all patterns in data will be revealed in a STAD graph. In particular, prior knowledge might reveal that for certain applications, specific datapoints should be pulled apart even if they are located close together when considering the complete high-dimensional space. We propose an extension of the STAD base algorithm to highlight other signals in the data by the inclusion of functions that acts as filters on data projections. 

Filters are an ordered set of values associated with an indexed sequence of natural numbers. They provide limits and thus a context to an arbitrary metric space. Filters can be defined from derived dimensions through statistical functions, subsets of dimensions or external data. Real sequences may be discretized by defining equidistant intervals based on scale or density. The addition of filters focuses on the exploration of data allowing, for example, to integrate domain knowledge. Formally, filters are defined as a space $Z$ with $n$ elements and $p$ dimensions in ${\rm I\!R}^p$ where a mapping exists between $X$ and $Z$, i.e., there is a function $f:X \rightarrow Z$. Filter functions theory is also present in topological methods \cite{bourbaki2013general} and the same filters used in TDA can be applied in STAD, providing in some cases similar shapes. However, although both share similarities conceptually, the fundamentals are different. While STAD aims to accentuate particular traits in the projection of a non-continuous set of points through filter functions, in TDA the filters are the basis of the projection itself and are used as an instrument to generate the manifold.

The inclusion of a filter replaces the first two steps of the STAD base algorithm (\ref{stad_base_step_1} and \ref{stad_base_step_2}) by a new approach composed by three steps: define the filter (Fig. \ref{fig:stad_lens}a), create a reduced distance matrix based on this filter (Fig. \ref{fig:stad_lens}b), and build a MST from the semi-complete weighted graph (Fig. \ref{fig:stad_lens}c). 

\subsubsection{Define the filter function}

Filter definition depends on data characteristics and the purpose of the analysis. As with other topological techniques, density estimates, centrality functions, orthogonal coordinates, a subset of dimensions and intermediate algorithmic results [\citenum{DBLP:journals/corr/abs-1811-00852}, \citenum{carlsson2012topological}] are also supported in STAD. The inclusion of filters aims to enrich data exploration through explicit prior knowledge \cite{wang2009defining} rather than hypothesis-free research. In practice, subsets of dimensions or external data tend to be most interpretable.

Filters in STAD are understood as a linear space where data follows a sequential order. However, the nature of the data included in a filter $Z$ can be diverse, and both the filter definition and interpretation must be adapted to it. For example, cyclical patterns in temporal data are common such as the day of the week or the month in a year. The last element of this cyclical pattern is as far from the previous as it is from the following although the sequential labeling does not reflect this, i.e., Sunday (day 7) is close to Monday (day 1) as is December (month 12) to January (month 1). In these cases, filters $Z$ must be represented in a polar space where the repetitive pattern is preserved \cite{weisstein2005spherical}.

Filters are mostly defined as one-dimensional or two-dimensional. Higher dimensionality although possible tends to over-restrict the space. When filters are real, a discretization process is required to transform them into a natural sequence of indices. In this paper, we present the real filter transformation by specifying $r$ as the number of intervals to divide each dimension in. Fig. \ref{fig:stad_lens}a illustrates the transformation of a real filter into a natural sequence. The effect of variations in the value of $r$ during the transformation influences the final network. The implications are discussed in section \ref{stad_interpretation}. The value $r$ can take independent values for each dimension when the filter dimension is greater than one, but the intervals must allow forming a single connected component network as STAD output. Empty intervals in a one-dimensional filter are omitted and the adjacency of the intervals is considered to the closest non-empty range. In filters of higher dimensionality, empty intervals are evaluated together with their neighbors defining a consistent grid. The algorithm to generate a consistent grid in STAD is provided as supplemental material.  

\subsubsection{Create reduced distance matrix based on filter}
\label{create_reduced_distance}

The inclusion of a filter $Z$ establishes limits in the metric. We can use these boundaries to introduce the effect of the filter in STAD by reducing the distance matrix $D_{X}$ and in consequence the complete graph $G_{X}$. We define three types of possible connections between datapoints (Fig. \ref{fig:stad_lens}a):

\begin{itemize}
\item \textit{Intra-edges} are all connections $e_{ij}$ where $i$ and $j$ belong to the same index.
\item \textit{Inter-adjacents} are all connections $e_{ij}$ where $i$ and $j$ belong to adjacent indices.
\item \textit{Inter-nonadjacents} are all connections $e_{ij}$ where $i$ or $j$ belong to different, non-adjacent indices. 
\end{itemize}

Based on these definitions, the distance matrix $D_{X}$ is reduced to $D_{X}^*$ by removing all non-adjacent connections (Fig. \ref{fig:stad_lens}b). The STAD process uses the distance matrix $D_{X}^*$ as input in the estimation of the filter. The derived graph from the distances becomes a semi-complete weighted graph $G_{X}^{*}$ where only links within and between adjacent intervals are considered. The reduction of connections draws networks based on the structure of the filter highlighting  properties of data such as the temporality of time-series or abnormality level of a centrality measure. Additionally, the performance of STAD with filters improves due to the smaller size of the distance matrix to be evaluated.

\subsubsection{Build MST from semi-complete weighted graph}

From $G_{X}^*$ the MST can be computed as described in \ref{stad_base_step_2}. However, one might want to ensure certain datapoints to be close together based on specific domain knowledge, even if they are further apart in high-dimensional space (or vice versa). In case the specific domain knowledge is expressed in a particular dimension, this would mean that the datapoints are far apart when considering all dimensions, but close together in the dimension under consideration.

Although the classical MST provides valid results in STAD with filters, we propose a version of the MST which better preserves the filter structure by prioritizing intra-edges in the process. Artificial connections (i.e. connections made as an artifact of splitting the data along the filters) are detected through community detection and re-evaluated globally. This process is split into these three steps:

\begin{enumerate}

\item The $MST_{intra}$ is created first inside of each index (intra-edge connections). Inter-adjacent connections are added after the $MST_{intra}$ computation to define a single connected component (Fig. \ref{fig:stad_lens}c left). 

\item The intra-edge connections from $MST_{intra}$ are evaluated through community detection using the original distances as weights. We implemented the random walk methodology Walktrap \cite{pons2005computing} due to its adaptability to short sequences. This step aims to detect distant points in high-dimensional space that were connected inside of each index. A sensitive configuration of community detection is desired to detect the different signals of data, as false negative divisions are automatically corrected in the following steps. All intra-edges falling in the same community and index are preserved and fixed. Remaining edges, i.e. discrepant intra-edge and inter-adjacent edges are omitted and re-evaluated in the following step (Fig. \ref{fig:stad_lens}c center).
 
\item Edges from the previous steps are preserved and act as a base, and additional edges are added until a single connected component is created (Fig. \ref{fig:stad_lens}c right). 
\end{enumerate}

\subsection{STAD network interpretation}
\label{stad_interpretation}

\begin{figure}[!t]
  \centering
  \includegraphics[width=\linewidth]{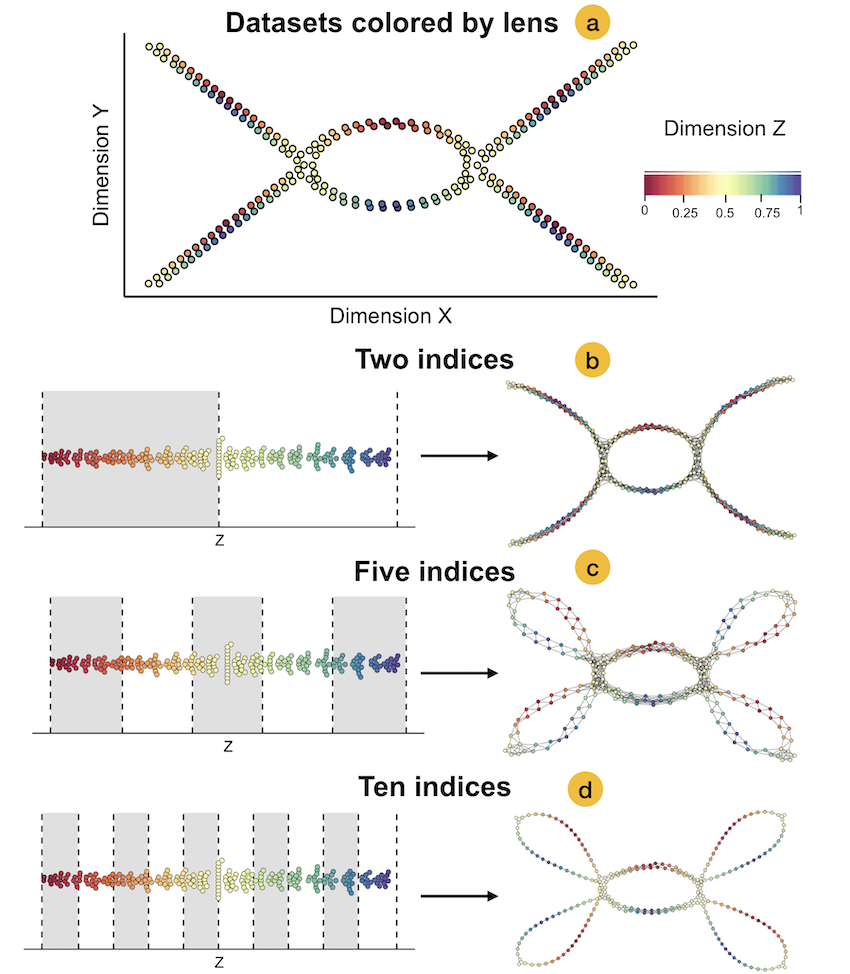}
  \caption{
  Effect of the filter and comparison of the number of indices. 
  (a) Three-dimensional dataset (spatial dimensions X and Y, and color dimension Z).
  (b) STAD network using dimensions X and Y as input and dimension Z (color) as the filter. The real filter is transformed into two equidistant indices.
  (c) Similar to (b) but dimension Z is transformed into five equidistant indices.
  (d) Similar to (b) but dimension Z is transformed into ten equidistant indices.
}
	\label{fig:stad_indices}
\end{figure}

STAD networks generate shapes which provide both global and local intuition of a data structure. Locals signals can refer to clusters, i.e., a homogeneous group of data points according to their similarity, but also broader meanings, for instance, a set of points with gradual dissimilarity (which presents itself as a flare). The graph density provides a notion of data distribution; homogeneous elements appear highly interconnected in the graph and dissimilar elements appear in non-adjacent sections of the graph. The visual edge length of STAD graphs indicates the similarity between the two vertices (see \ref{node_link_diagram}). Specific graph layout algorithms such as the force-directed layout will search for an equilibrium between these edge lengths and their optimization function, e.g., to minimize overlap of nodes and/or edges \cite{hua2018graph}.

The inclusion of filters intensifies specific information contained in their dimensions. Fig. \ref{fig:stad_indices} exposes the effect of filters in a comparable setting where the same variable has been split in a different number of natural indices. According to the distance matrix detailed in \ref{create_reduced_distance}, when the number of filter indices is two or smaller, no non-adjacent connections exist and therefore we obtain a STAD network identical to the filter-free approach (Fig. \ref{fig:stad_indices}b). A higher number of indices produces a fine-grained representation of the filter definition but penalize the structural representation of points contained in the underlying dataset $X$. If the number of indices in the filter is equivalent to the number of elements in the dataset, the generated network is forced to connect the adjacent indices. In this case, the STAD network exposes the structure of the filter instead of the structure of data. Additional features of the graph (e.g., node color and/or size as used below) can aid in the interpretation of the data.

\section{Case Studies}
\label{case_studies}

We applied STAD to two real-world datasets. We present results, derived insights extracted from the shapes and discuss choices on the filter selection. The visual analytics approach helps to discover non-evident patterns in data through the connection between the points describing data shapes as flares and loops. 

\begin{figure*}
  \centering
  \includegraphics[width=\linewidth]{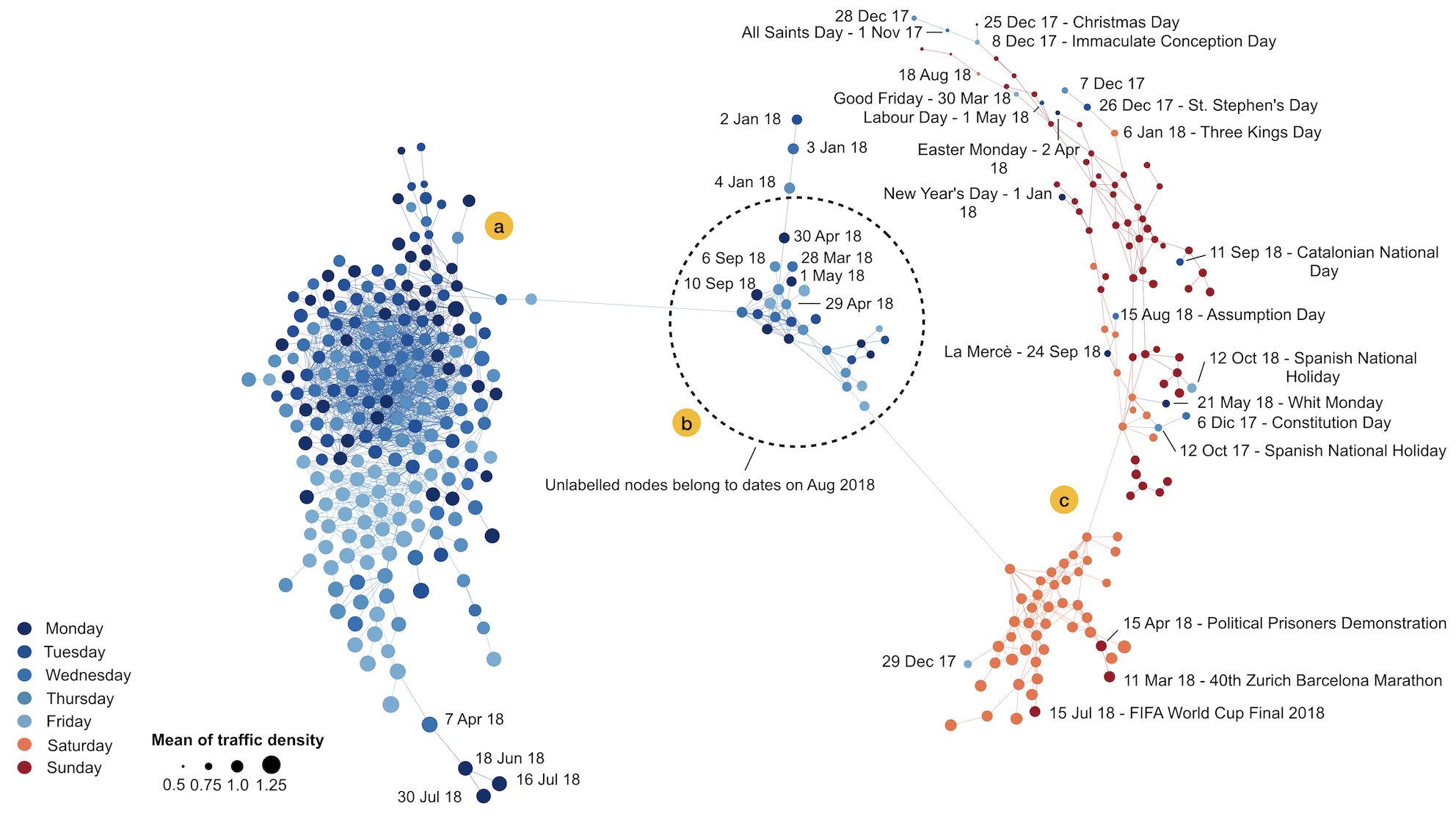}
  \caption{
  STAD network for the characterization of traffic in Barcelona from October 2017 to November 2018, reflecting the differences between workdays, weekends, local holidays and vacation periods. The network is visualized using the ForceAtlas2 layout.
  Each node represents the temporal activity of traffic of a single day and is linked to other days with similar behavior. Color corresponds to the day of the week and size to the mean of the traffic density. (a) Group of workdays. (b) Workdays during holiday period. (c) Weekends and official holidays. 
}
	\label{fig:bcn_nolens}
\end{figure*}

\subsection{Barcelona traffic}

We collected a dataset from the public repository Open Data BCN \cite{WinNT} which contains traffic activity in the city of Barcelona. The analysis was performed with all records from October 2017 until November 2018. The dataset describes measurements of traffic density collected every five minutes in 534 locations of the city which is stored as an ordinal variable from one to six, one corresponding to freely moving traffic and six to stand-still. We explored the daily changes in the city by averaging the individual sections into a one-dimensional time series for each day. Similarity between days was computed using Euclidean distance to identify differences at identical timestamps. In this section, we will discuss two analyses: one without filter and one using a two-dimensional filter composed of the week number and the daily mean of the densest point in the city.

\subsubsection{Filter-free STAD analysis of Barcelona traffic}

Filter-free STAD analysis results are shown in Fig. \ref{fig:bcn_nolens} where we identify three different patterns. In this example, nodes are colored by day of the week (blue shades correspond to workdays from Monday to Friday, orange refers to Saturday, and red to Sunday) and the size of the node indicates the mean of traffic density in that day. The most significant signals correspond to the difference in activity between weekdays (Fig. \ref{fig:bcn_nolens}a) and weekends (Fig. \ref{fig:bcn_nolens}c). The groups of workdays on the left are highly connected indicating the high similarity between these days. In the center of the graph (Fig. \ref{fig:bcn_nolens}b), we find a subset of days between the largest group of workdays and weekends. This subset corresponds to low activity days in the city; more specifically, they are workdays in the first week of January, Easter week and the month of August. These periods of the year traditionally are associated with holiday periods and are distinguishable from the rest of patterns in data. On the right of the graph, we recognize the weekend and official holidays. Inside this sub-network, we recognize two more groups which mostly correspond to the two days of the weekend. Saturdays are days with higher activity than Sundays as reflected on the node size of the figure. Official holidays behave like a typical Sunday; we highlight Christmas day as the day of the year with the smallest traffic activity, located at the extreme top-right. In contrast, there are some Sundays with higher traffic activity which have been related to some featured event. For example, we name the Political Prisoners Demonstration on April 15th, the 40th Zurich Barcelona Marathon on March 11th and the Final World Cup 2018 on July 15. These days are associated with a higher movement of people and the closing of some section of the city.

Although the connectivity of the network does not provide additional structural insights, the color of nodes helps to recognize weaker signals in the graph as well. More concretely, we can see that Fridays are particularly clustered. Digging deeper into the data we can identify a peak of activity between 14:00 and 16:00 on Fridays (see image in supplemental material). The increased traffic is associated with departures leaving the city.  

The global structure of the network presents coherent connectivity between the groups according to their traffic density: the group with highest traffic congestion i.e. typical workdays (Fig. \ref{fig:bcn_nolens}a) connects to the group with workdays on holiday period (Fig. \ref{fig:bcn_nolens}b) and this is linked to the weekends days (Fig. \ref{fig:bcn_nolens}c).

\subsubsection{STAD analysis of Barcelona traffic using two-dimensional filter}

\begin{figure*}
  \centering
  \includegraphics[width=\linewidth]{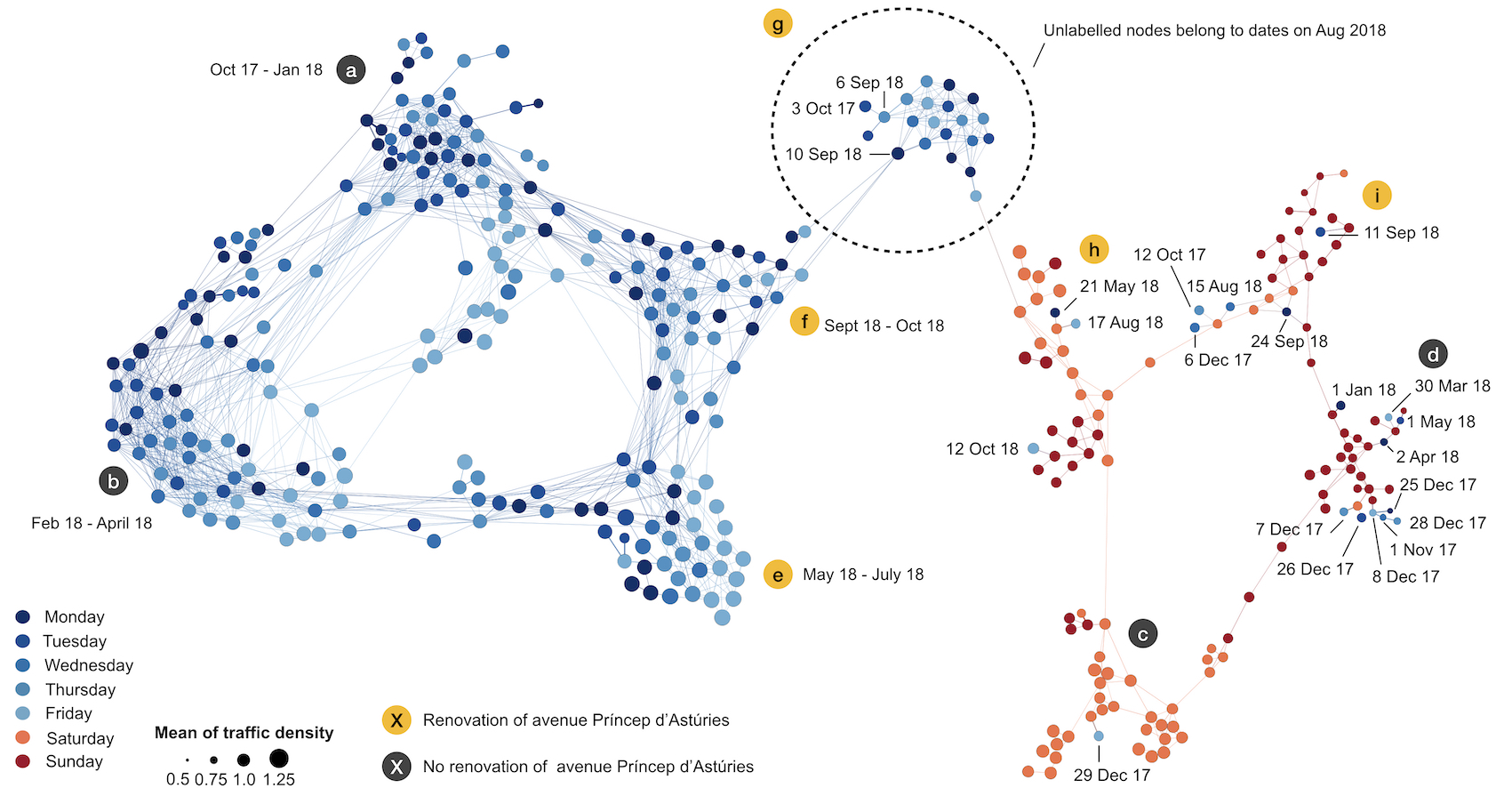}
  \caption{ STAD network using the week-number and the mean of the densest point on Barcelona traffic. ForceAtlas2 was the selected layout algorithm, and MST was built as described in \ref{stad_base_step_2}.
  The two loops indicate the differences in traffic activity during the year and renovations performed in a popular avenue of the city which caused the closing of this part. The visual clusters of the network are identified with colors to indicate if they belong to the renovation period: (a-d) No renovation of avenue Princep d' Asturies, (e-i) Rennovation of avenue Princep d' Asturies.
}
	\label{fig:bcn_lens}
\end{figure*}

We continue the analysis of Barcelona traffic by incorporating filter functions to identify additional signals. We applied STAD using a two-dimensional filter composed of the week number and the mean of the densest point in the city for each day. The resulting network in Fig. \ref{fig:bcn_lens} maintains the three groups from the approach without filters (Fig. \ref{fig:bcn_nolens}) but additional features are revealed. Two additional loops are present, one in the group of workdays and the other on weekends. Further investigation indicated that these structures correspond to renovation works \cite{renoWork} starting in May 2018 which resulted in the closing of a transversal avenue in the west of the city (Fig. \ref{fig:bcn_lens}e-i).

The visual gaps in the graph, for instance, between groups a and b, are created by public or bank holidays, which end up in the cluster of the weekend days (groups c-d and g) and generate this weaker connectivity between the indices of the temporal filter week number. Likewise, the gap between groups e and f is due to August present at the center of the graph. The circular pattern of traffic between years is reflected in the connectivity between groups f and a. In the weekends we can identify the same separation due to the renovation works (c-d vs h-i). 

\subsection{Air-quality in Castile and Le\'on }

\begin{figure*}[!ht]
  \centering
  \includegraphics[width=\linewidth]{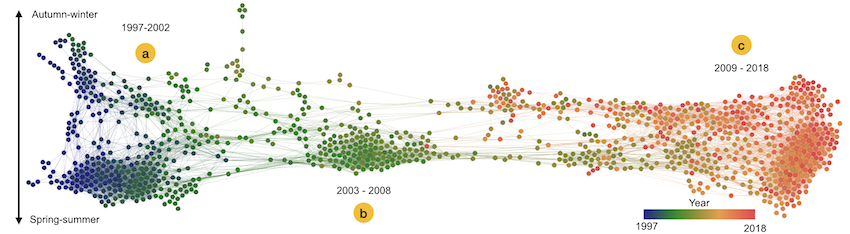}
  \caption{Filter-free STAD network describing the evolution of air-quality from 1997 to 2018. The chemicals measured were: CO, NO, NO2, O3, PM10, PM25, and SO2. The graph displays a clear evolution of air quality identifying three different periods: (a) Between 1997 and 2002 autumn-winter and spring-summer follow different patterns due to an increase of CO, NO, NO2, and SO2 in autumn-winter. (b) The concentration of PM10 and PM2.5 decreased in 2003 compared to 2002. (c) High connectivity of nodes reveals smaller variations of NO, NO2, and CO values across the year.}
	\label{fig:air_quality_nolens}
\end{figure*}

We applied STAD on the air-quality dataset collected from the Castile and Le\'on initiative in Spain \cite{opendataair} to illustrate STAD as a visualization tool for the identification of patterns on high-dimensional time series. The dataset contains daily measurements of seven chemicals such as carbon monoxide (CO), nitrogen oxide (NO), nitrogen dioxide (NO2), ozone (O3), sulfur dioxide (SO2) and particulate matter 10 and  2.5 (PM10 and PM2.5). The measurements have been collected at different locations from January 1997 to June 2018. The data was aggregated by week due to the presence of missing values. The explored multidimensional time series contains 1139 records with seven dimensions, and we computed the Euclidean distance to evaluate the similarity between elements. 
The resulting STAD network graph is presented at Fig. \ref{fig:air_quality_nolens}. The structural shape generates an intrinsic separation of time identifying changes in the air-quality over the years. We recognize three dense groups of points which mainly correspond to different periods: 1997-2002 (Fig. \ref{fig:air_quality_nolens}a),  2003-2008 (Fig. \ref{fig:air_quality_nolens}b) and 2009-2018 (Fig. \ref{fig:air_quality_nolens}c). These visual splits identify relevant changes in the air-quality. The vertical position of nodes provides an intuition of seasonality, i.e. nodes on top of the network correspond to autumn-winter dates and nodes on the bottom to spring-summer. The coloring of nodes by seasonality is provided as supplemental material. 

To investigate seasonality signals further, we extend our exploration by incorporating the week number as the filter in STAD (Fig. \ref{fig:teaser}). The network conserves the signals identified in Fig. \ref{fig:air_quality_nolens} although they also reveal additional ones. For instance, between 1997 and 2002 (Fig. \ref{fig:teaser}a) two groups are evident, corresponding to different seasons (autumn-winter and spring-summer). In contrast, between 2009 and 2018 (Fig. \ref{fig:teaser}c) the nodes are highly connected, forming a cycle. Further analysis on the network shapes indicates the following: 
 
 \begin{figure*}
  \centering
  \includegraphics[width=\linewidth]{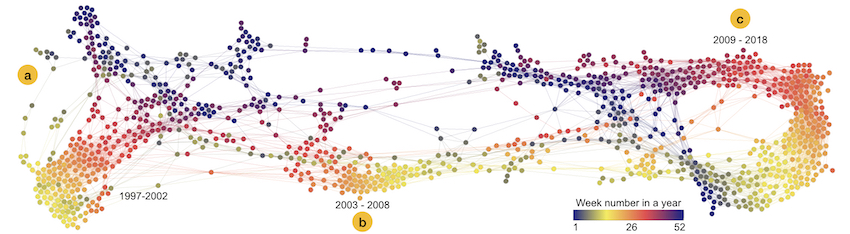}
  \caption{ STAD network using week number as the filter to emphasize distinctive periods of time. ForceAtlas2 was used to draw the network, and MST was created following section \ref{stad_base_step_2}. The three periods are: a) period 1997-2002, (b) period 2003-2008 and, (c) period 2009-2018. }
	\label{fig:teaser}
 \end{figure*}

\begin{itemize}
\item The two different groups identified at (Fig. \ref{fig:teaser}a) are mainly caused by the chemicals nitrogen oxide and dioxide, carbon monoxide, and sulfur dioxide. These measurements are higher during the autumn-winter and are related to the burning of fossil fuels \cite{weiss1976production}. In recent years, electric systems started to substitute the previous technologies \cite{carbonHeating}. This fact is visible in the period 2009 to 2018 (Fig. \ref{fig:teaser}c) where the difference between seasons is less evident.

\item The gap between 2002 and 2003 at spring-summer (Fig. \ref{fig:teaser}b) indicates the decrease of particulate matter PM10 and PM2.5, which is related to vehicle emissions \cite{rodriguez2004comparative}. This period corresponds to new vehicle restrictions \cite{tzamkiozis2010diesel}.

\item Different coloring of nodes may help reveal additional patterns in data (see images in supplemental material). For example, ozone fluctuates according to the season period. During the spring-summer, ozone levels are higher due to variations in sunlight and UV radiation \cite{scheel1997spatial}. In addition, concentration of carbon monoxide, nitrogen, ozone, sulfur dioxide, and particulate matter decreases gradually over the years, reaching stability in 2010. This finding is associated with an improvement of air-quality \cite{vedrenne2015integrated}.

\end{itemize}

\section{Discussion}

In this section, we discuss some limitations of the STAD methodology, their possible solutions and open challenges that remain to be addressed. 

\subsection{Scalability}

While STAD analysis of some datasets results in sparse networks with easily interpretable structures, other datasets end up represented in more complicated networks. As the algorithm works at the level of the individual datapoints, the analysis of large datasets comes at a significant computational cost. In addition, drawing of a large resultant network also becomes cumbersome. For example, the construction of the largest network (air-quality dataset) in this paper takes 1.2 minutes. All code was implemented in R with a single-thread and run on an Apple laptop MacBook Pro (dual-core, Mid 2014).

\subsubsection{Computational scalability}
\label{comp_scalability}
In STAD, the recursive computation of distances in the unit-distance graph comprises the main bottleneck of network estimation. A possible approach to alleviate this issue is to work with a smaller sample of the initial dataset. Preliminary tests have indicated that such smaller dataset retains the same visual structure as the full-size dataset, while not suffering from the high computation cost. 
The addition of edges upon the MST is a cumulative process (\ref{add_edges}), i.e., if an edge $i$ with weight $w_{i}$ - with larger weight meaning larger dissimilarity - is added into the network all edges with smaller weight are part of the network $U_{i}$. When the algorithm determines the optimal network, it finds an edge-weight threshold which establishes the resulting number of connections. This optimum can be calculated on a subsample of the full-size dataset. Multiple iterations on a down-sampled dataset can be performed in a parallelized setting providing a more robust threshold estimation.

\subsubsection{Visual scalability}

When large networks are considered, the number of links might become an issue for visualization, resulting in the dreaded "hairball". Current approaches on graph layouts \cite{jacomy2014forceatlas2} could manage up to a million nodes if the network is sparse [\citenum{gomez2018visualizing}, \citenum{hua2018graph}]. Consequently, additional transformations such as aggregation to reduce the number of nodes and/or edges are still needed. A possible solution can be found in community detection pipelines \cite{fortunato2016community}, such as MCLEAN \cite{alcaide2018mclean}, that simplify this visual representation.

\subsection{Addition of edges}

As mentioned in \ref{comp_scalability}, the addition of edges from the MST follows a sequential and cumulative procedure based on their distances. The incremental approach may cause edge redundancy and contribute to increasing clutter in the plot. A non-additive and refined procedure such as an evolutionary algorithms \cite{back2018evolutionary} might reduce the number of links conserving the association with the original distance matrix $D_{X}$. Nevertheless, the use of evolutionary algorithm would also penalize the performance on the network estimation due to the assessment of crossovers in every iteration. Moreover, elimination of unnecessary edges does not change the structure of the resultant network, and such approach does not guarantee a benefit in making new features recognizable.

\subsection{Stability and reproducibility}

The optimal number of links is dictated by the association between $D_{X}$ and $ D_{U}$. The SA algorithm estimates an optimum through a stochastic procedure which can generate different results in each iteration. However, the correlation curve describes a soft convexity with a quasi-constant function around the maximum.

Filters can bring additional signals in the underlying structure of data to the foreground. In extreme cases with very small bins, these might however fragmentize the original structure, although these changes have demonstrated to be reasonably robust (Fig. \ref{fig:stad_indices}).

\subsection{Comparison with related techniques}

\begin{figure*}
  \centering
  \includegraphics[width=\linewidth]{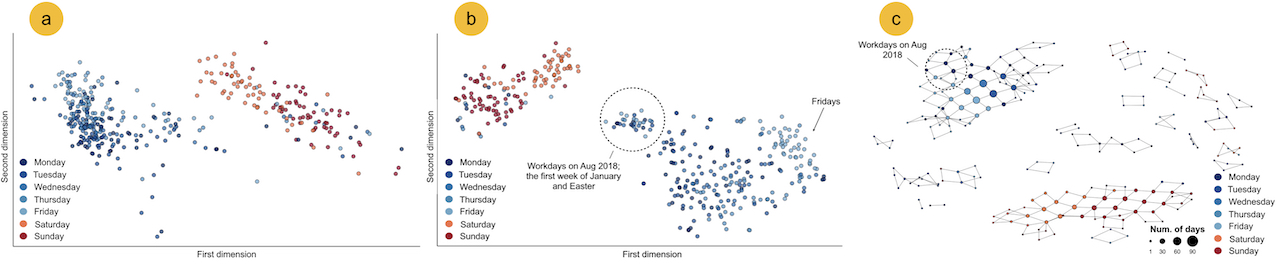}
  \caption{ Other methods for Barcelona traffic data. (a) NMDS projection. (b) t-SNE projection on Barcelona traffic. Perplexity = 30 and 500 iterations. (c) The Mapper algorithm. Lenses: two-dimensional NMDS with 15 intervals and 50\% overlap.
}
	\label{fig:bcn_evaluation}
\end{figure*}

Alternative methodologies can reveal equivalent signals in data to STAD, especially in the filter-free approach. In this section, we present the different results between STAD and related techniques in the Barcelona traffic case. Non-linear multidimensional scaling (NMDS), t-SNE, the Mapper algorithm, and hierarchical clustering were selected due to the possibility of establishing distances matrices as inputs. Fig. \ref{fig:bcn_evaluation} presents the resulting projections for each method:

\begin{itemize}
\item NMDS (Fig. \ref{fig:bcn_evaluation}a) captures the difference between weekdays and weekends, although low activity days are not clearly separated from the rest of workdays.
\item t-SNE identifies the three most relevant patterns presented with STAD (Fig. \ref{fig:bcn_evaluation}b). Fridays are also distinguishable in the embedding. However, densities are poorly preserved. For instance, workdays (especially from Monday to Thursday) have low variability in the dataset but the projection generates an excessive separation between them. 
\item The Mapper algorithm (Fig. \ref{fig:bcn_evaluation}c) requires the definition of lenses. We selected the projection of NMDS to have a comparable view with the evaluated methods. The network presents a summarization of NMDS, but no additional signal can be detected. 
\end{itemize}

\section{Conclusion and future work}

STAD propose a parameter-free methodology to visualize the structure of high-dimensional datasets, allowing for the identification of signals by means of geometrical shapes as flares and loops. Edges in the graph correspond to the similarity between datapoints; therefore, similarities between individual datapoints are used to encode the higher-level patterns in the resultant graph and the resulting visualization is a compressed projection of the distance matrix into a free-scale space. In addition, integrating filters adds an additional perspective to the exploratory analysis.

On R implementation is available at \url{https://github.com/vda-lab/stad}; Python and Clojure implementations are under development. Results presented in this paper have been generated with this R implementation. The final graphs included in section \ref{case_studies} were enhanced through Gephi \cite{bastian2009gephi}. 

Future work includes improving the efficiency of computational methods by retaining the information from previous iterations and approximating the shortest path \cite{holzer2014approximation}. In addition, we also aim to devise novel visual approaches to compare and interpret networks structures in an integrated environment. 

\ifCLASSOPTIONcompsoc
  \section*{Acknowledgments}
\else
  \section*{Acknowledgment}
  
\fi

The authors wish to thank Danai Kafetzaki for proofreading. This work was supported in part by the IWT/SBO 150056 project ”ACquiring CrUcial Medical information Using LAnguage TEchnology” (ACCUMULATE).

\ifCLASSOPTIONcaptionsoff
  \newpage
\fi



\bibliographystyle{IEEEtran}
\bibliography{template}
\end{document}